\documentclass[preprint,12pt]{elsarticle}

\usepackage{graphics}

\usepackage{amssymb}

\journal{Physica A}

\begin{document}

\begin{frontmatter}


\author[a]{C.~Seidel}
\ead{seidel@mpikg.mpg.de}
\author[b]{Yu.A.~Budkov}
\ead{urabudkov@rambler.ru}
\author[c]{N.V.~Brilliantov\corref{cor1}}
\ead{nb144@leicester.ac.uk}
\address[a]{~Max Planck Institute of Colloids and Interfaces, Science Park Golm, D-14424
Potsdam, Germany}
\address[b]{Institute of Solution Chemistry, Russian Academy of Sciences, 153045 Ivanovo, Russia}
\address[c]{Department of Mathematics, University of Leicester, Leicester LE1 7RH,
United Kingdom}

\cortext[cor1]{Corresponding author}

\title{Field-regulated force by grafted polyelectrolytes}




\begin{abstract}

 Generation of mechanical force regulated by external electric field is studied
both theoretically and by molecular dynamics (MD) simulations. The force arises in deformable bodies
linked to the free end of a grafted polyelectrolyte chain which is exposed to electric field that favours its
adsorption. We consider a few target bodies with different force-deformation relations
including (i) linear and (ii) cubic dependences
as well as (iii) Hertzian-like force. Such force-deformation relations
mimic the behaviour of (i) coiled and (ii) stretched polymer chains, respectively, or (iii) that of a
squeezed colloidal particle. The magnitude of the arising force varies over a  wide interval although the
electric field alters within a relatively narrow range only. The predictions of our theory agree
quantitatively  well with the results of numerical simulations. Both cases of zero and finite electrical
current are investigated and we do not obtain substantial differences in the force generated. The 
phenomenon studied could possibly be utilised to design, e.g.,  vice-like devices to fix nano-sized objects.
\end{abstract}


\end{frontmatter}


\section{Introduction}

During last decades investigations of the response of charged polymers, so-called polyelectrolytes, to
external electric field  attracted much scientific attention.
\cite{Muthu1987,Bajpai1997,Borisov1994,Boru98,Joanny98,Muthu2004,Dobry2000,Dobry2001,Borisov2001,
Netz2003,Netz2003a,Borisov2003} In particular, the application of novel  experimental techniques such as
atomic force microscopy, which can explore the behaviour of a single chain,  provided new insight
into the problem.\cite{FriedsamGaubNetz2005} If the free end of a grafted polyelectrolyte is under load,
while the chain itself is exposed to electric field that   favours its adsorption, the configuration of
the chain is determined by both the field and the force applied to the free end.\cite{Joanny98}
In particular, the force may arise if the chain is mechanically coupled to a deformable, nano-sized
object, such as another polymer chain or a colloidal particle. Any
deformation of the object (target body) gives  rise  to a restoring force (see Fig. 1). The
force and the length of the bulk polymer segment, i.e., the part pulled off from the adsorption layer,  are determined in a self-regulated
manner:~The larger the field  the stronger the attraction of the polyelectrolyte to the surface and the
shorter becomes the bulk piece of the chain. Due to the mechanical coupling, however, shrinking of the bulk part induces
increasing deformation of the target body which results in an enlarged restoring  force. 
Therefore, varying the external field one can control chain configuration and generated force.
The latter crucially depends on the mechanical properties of the target body, i.e., on its response to
deformation. Qualitatively, one can have either linear response, like for a linear spring that models
small deformations of a coiled polymer, or nonlinear one, as for strongly stretched chains or squeezed
colloidal particles, see Fig.~1 and Fig.~2.

Such polyelectolyte-based setups can be utilised to generate mechanical force regulated by external
electrical field. There are numerous possible applications of the mechanism, ranging from nano-scale
devices, designed to fix nano-sized objects ("nano-vice"), up to much larger meso- or marco-scales
such as artificial muscles.

%
From a fundamental point of view, as well as with respect to possible applications, two questions are of
a primary interest:~(i) How does the force generated depend on field strength and elastic properties
of the target body? (ii) How does the chain configuration alter with varying electric field? The analysis
of these problems may be also important in the context of the efficiency of future nano-devices such as a vice-like one suitable to fix nano-sized objects

In our model the non-anchored end of a grafted polyelectrolyte chain is linked
to a deformable target body that is  modelled by various types of springs both linear and non-linear ones.
Perpendicular to the grafting plane a variable electric field is applied that favours polyelectrolyte adsorption. Moreover the field drives counterions away from the polyelectrolyte up  to the top boundary. As
a result, at zero electrical current the counterion subsystem is practically decoupled from
polyelectrolyte provided the specific volume per chain is large enough.

In a previous study~\cite{BrilliantovSeidel2012}  we investigated theoretically as well as by means of
molecular dynamics (MD)  the effect of a constant force which pulls the polyelectrolyte chain away from
the adsorption plane. We analysed the case of zero electrical current and found a very good agreement
between theoretical prediction and simulation results.  There, we presented also first simulation
results on force generation for chains linked to a target body. For that problem, however, no analytical theory has been given.
%


In the present paper we report both extensive MD simulation results  and a theoretical study of the
generation of mechanical force by external electrical field.  A few different force-deformation
relations of the target body have been explored. In particular, we analyse linear, non-linear (cubic)
and Hertzian springs, that mimic coiled Gaussian chains, strongly stretched chains and squeezed
colloidal particles, respectively. We present a analytical theory which is in a good agreement with the
simulation results. In addition to the studies with zero current at equilibrium, we performed
simulations where a finite current across the box is allowed and found that the impact of current on
chain configuration and response force is negligible for the parameter range addressed.

The paper is organised as follows: In the next Section, we give details of the numerical setup and
present simulation results. The analytical theory is introduced in Section 3 and subsequently we
compare our theoretical predictions with simulation data. Finally, in Section 5 we summarise our findings.

\section{MD simulations of a grafted polyelectrolyte in electric field}
%
In the simulations reported the polyelectrolyte is modelled by a freely jointed bead-spring chain of
length $N_{0}+1$ which is anchored by one uncharged end-bead to a planar surface at $z=0$. All the
remaining $N_{0}$ beads carry one (negative) elementary charge. To satisfy electroneutrality, $N_{0}$
monovalent free counterions of opposite charge are added. For simplicity, they are treated as particles
of the same size as monomers. The implicit solvent is assumed to be a good one modelled by a short-ranged, purely
repulsive interaction between all particles. We describe it by a shifted Lennard-Jones
potential. Along the chain, neighbouring beads are connected by a finitely extensible, nonlinear elastic
FENE potential. With our choice of parameters, at zero force we obtain a bond length $b \simeq
\sigma_{LJ}$ where $\sigma_{LJ}$ is the Lennard-Jones parameter. All particles except the anchor bead
are exposed to a short-ranged repulsive interaction with the grafting plane at $z=0$ and with the upper
boundary at $z=L_{z}$. All the charged entities interact with the bare Coulomb potential the strength of
which is fixed by the Bjerrum length $l_B=e^2/\varepsilon k_B T$, where $e>0$ is the elementary charge,
$\varepsilon$ is the dielectric constant of the medium, $T$ is temperature and $k_B$ is the Boltzmann
constant. In the simulations we set $l_B=\sigma_{LJ}$ and use a Langevin thermostat to hold the
temperature $k_{B}T=\epsilon_{LJ}$ with $\epsilon_{LJ}$ being the Lennard-Jones energy parameter.
Details of simulation model and method can be found elsewhere \cite{CSA00,KUM05,KUM07}.

The free chain end is linked to a target body that is modelled by springs of different force-deformation
relations: (i) linear spring, $f = \kappa (\xi- h_0)$, where $f$ is the response force, $\kappa$  is the
strength of the spring, $\xi$ and $h_0$ are the lengths of deformed and undeformed spring, respectively.
Such a force-deformation relation mimics, e.~g., the behaviour of a polymer chain in a coiled Gaussian
state. (ii) nonlinear spring, $f = \kappa (\xi- h_0)^{\gamma}$, with $\gamma = 3$, which models a
stiffer body, e.~g., a chain in an intermediate state between the coiled and stretched one. (iii)
Hertzian force $f = \kappa (h_0 - \xi)^{3/2} \theta(h_0-\xi)$, which corresponds to the
force-deformation relation for a squeezed colloidal particle of a diameter
$d_c=h_0$.\cite{Hisao:2009,Hisao:2010} Note that there are two different setups where the spring is
anchored either A) at the top wall (see Fig. 1, left panels) or B) at the grafting plane (see
Fig.1, right panels and Fig.2. In our model, we assume that (i) the anchor of a
spring is fixed and (ii) the spring is oriented perpendicular to the wall. Hence, the instantaneous
length of the spring is given by A) $\xi = L_z - z_N$ or B) $\xi = z_N$, respectively, where $L_z$ is
the box height in $z$-direction and $z_N$ is the $z$-coordinate of the end segment of the
polyelectrolyte linked to the target body.

In this paper, we report simulation results obtained for polyelectrolytes of total  chain lengths
$N_{0}=$ 320. The footprint of the simulation box is $L_{x}\times L_{y}$ = 424 $\times$ 424 (in units of
$\sigma_{\rm LJ}$). The box height is $L_{z}$ = 160, i.e., both the effective surface charge density of
grafting plane and box height are identical to those chosen in our previous
study.\cite{BrilliantovSeidel2012} Note that previously we studied equilibrium properties but
neglected any current across the box which could, however, reduce the force generated and limit the
efficiency of possible nano-devices. In the present study, in addition to equilibrium properties we
checked the effect of a finite current which is treated by the following simple model: counterions
coming close to the top boundary are removed with a finite probability $p_c$ and reinserted just above
the bottom boundary. The average current density $j $ in this model is linearly related to the
probability $p_c$ as $j = a\, p_c$, where the constant $a$ may be expressed in terms of the average
counterion density in the top layer and the simulation time-step.\footnote{The average current density
reads $j= e\, \Delta l  n p_c (\Delta t_{\rm sim})^{-1}$, where $\Delta l$ is the thickness of the top
layer from which counterions are removed, $n $ is the average density of counterions in this layer, and
$\Delta t_{\rm sim}$ is the simulation time step. Hence the constant $a$ reads, $a=e\, \Delta l n(\Delta
t_{\rm sim})^{-1}$} For the purposes of the present study we do not need its value. Varying $p_c$
from zero (vanishing current) to a maximal  value of $p_c = 0.05$ we alter the electric current and cover all
possible bulk distributions of counterions that range from strong
accumulation near the upper plane (cathode) to homogeneous distributions
across the whole box height (see Fig.~3).
As shown below in detail (see Fig.~6), we find  that the influence of a finite current is
practically negligible. This behaviour occurs probably because of the very low bulk density of
counterions in our setup. Moreover, while for small current almost all counterions are accumulated in
the upper layer close to cathode leaving the polyelectrolyte unscreened, at  larger currents counterions
move rather fast and are not able to screen the polymer charge efficiently.  Although hydrodynamic
interactions are not present in the simulation model used, we expect that the basic behaviour would
quantitatively persist also in a refined dynamic model. As the result and in sharp contrast to the
field-free case,\cite{Brill98,Winkler98,Gole99,Diehl96,Pincus1998,MickaHolm1999,Naji:2005}  the
counterion subsystem is practically decoupled from the polyelectrolyte  which drastically simplifies the
theoretical  analysis.
%

%
A typical simulation snapshot at vanishing current is shown in Fig.~2. We find that already
at relatively weak fields, $Eqeb/k_BT \geq  0.1$, with $qe$ being the monomer charge, the adsorbed part of
the polyelectrolyte exhibits an almost two-dimensional structure, with small loops, rising  out of the
plane up to a hight of about a monomer radius. At the same time, the pulled-off bulk part is strongly
stretched along the direction of the force and the polymer bonds are almost perfectly aligned perpendicular to the grafting plane.
Simulation results are shown in Figs.~4 - 7 and discussed along with the
corresponding theoretical predictions in Sec. 4.

\section{Theory}
The total free energy of the system $F_{\rm tot}(N)$ can be written as a function of the number of desorbed
chain monomers in bulk, $N$. It consists of a few contributions
\begin{equation}
\label{eq:F_tot} F_{\rm tot}(N)= F_{\rm bulk}(N)+F_{\rm surf}(N_s)+U_{\rm spring},
\end{equation}
where $F_{\rm bulk}(N)$ is the free energy of the bulk part of the chain, i.e., of the desorbed part,
$F_{\rm surf}(N_s)$ is the free energy of the adsorbed part, which depends on the number of
adsorbed monomers $N_s=N_0-N$ and $U_{\rm spring}$ is the mechanical energy of the target body modelled by a spring.
The bulk free energy $F_{\rm bulk}(N)$ is calculated as the sum of two contributions, $F_{\rm bulk}=F^{E}_{\rm
bulk}+F^{\rm self}_{\rm bulk}$, where $F^{E}_{\rm bulk}(N)$ corresponds to the interaction of the desorbed part of the chain
with the external field $E$ and $F^{\rm self}_{\rm bulk}$ accounts for the self-interaction of the bulk part, i.e., for
the electrostatic interaction between charged monomers. For strongly stretched chains, i.e., chains under strong stress, the entropic component of the bulk part can be neglected and
$F^{E}_{\rm bulk}$  equals the energy of a charged chain in electric field
\begin{equation}
\label{eq:F_bulk0} F^{E}_{\rm bulk} \simeq \sum_{i=1}^N eq \varphi(z_i) = -Eeq\sum_{i=1}^N z_i \,  ,
\end{equation}
with $z_i>0$ being the distance of the $i$-th chain monomer from grafting plane and $eq$ is the monomer
charge. In Eq.~(\ref{eq:F_bulk0}), we ignore counterion screening of the chain as well as of the
boundary planes (which is justified for our setup, see the discussion above) and approximate the
electrostatic potential as $\varphi(z) =-Ez$, where $E$ is the (constant) electric field in
$z$-direction. For simplicity, we neglect bond stretching and assume that for strongly aligned chains
all bonds are directed in vertical direction (see Fig. 2. Then the height of the $i$-th
monomer can be written $z_i \simeq  i b$ which allows to sum up
\begin{equation}
\label{eq:z_k} \sum_{i=1}^N  z_i \simeq  \sum_{i=1}^N  i b =\frac{bN(N+1)}{2}
\end{equation}
and Eq.~(\ref{eq:F_bulk0}) yields
\begin{equation}
\label{eq:F_bulk} \beta F_{\rm bulk}^{E}\simeq -\frac{\beta qeEbN(N+1)}{2}=\frac{\tilde{E}N(N+1)}{2}\, ,
\end{equation}
with $\tilde{E}=\beta|q|eEb$ and $\beta=1/k_{\rm B}T$ where we take into account the negative sign of $q$ (for the assumed
positive field $E$).
For stretched chains, the self-interaction part $F^{\rm self}_{\rm bulk}(N)$ may be approximated by the electrostatic energy of $N$ charges uniformly placed on a linear string of length $z_N \simeq Nb$
\begin{eqnarray}
\label{eq:F_self_bulk}\beta F_{\rm bulk}^{\rm self}&\simeq& \sum_{i=1}^{N-1}\sum_{j=i+1}^{N}\frac{\beta
q^2e^2}{\varepsilon|j-i|b} \nonumber \\
&\simeq& q^2\tilde{l}_{B}\left[\ln\Gamma(N)+(N-2)\gamma_{E}\right],
\end{eqnarray}
where $\tilde{l}_{B}= (e^2/\varepsilon k_BT)/b=l_{B}/b $ is the reduced Bjerrum length, $\Gamma(x)$ and $\gamma_E \simeq 0.57721$ are
$\Gamma$-function and Euler constant, respectively.

In the present study we analyse the effect of different load modelled by linear and non-linear springs.
In general terms, we write
\begin{equation}
\label{eq:spring} \beta U_{\rm spring}=\frac{\beta
\kappa(l_{0}-z_N)^{\gamma+1}}{\gamma+1}=\frac{\tilde{\kappa}(\tilde{l}_{0}-N)^{\gamma+1}}{\gamma+1},
\end{equation}
where $l_0=L_z-h_0$ for setup A) (linear and non-linear springs) and $l_0=d_c$ for setup B) (Hertzian spring), $\tilde{\kappa}=\beta\kappa b^{\gamma+1}$ and $\tilde{l}_{0}={l}_{0}/b$. Furthermore, in the above
equation  we use again the relation $z_N \simeq Nb$. Applying the latter equation to the case of
Hertzian springs one has to add an additional factor $\theta (l_0-z_N)$ with $\theta(x)$ being the
Heaviside step-function, skipped here for brevity. This factor reflects the particular feature that
Hertzian springs respond only to compression. For a harmonic spring one has $\gamma=1$  and a linear
force-deformation relation.

The surface part of the free energy $F_{\rm surf}$ is written as a function of the
radius of gyration $R_g$ and the number of adsorbed monomers $N_s=N_0-N$
\begin{equation}
\label{eq:surface} \beta F_{\rm surf}(R_{g},N_s)=\alpha\frac{N_{s}^2q^2}{R_{g}}l_{B}-\log
\left[(2\pi)^{N_{s}}P_{N_{s}}(R_{g})\right]\, ,
\end{equation}
where the first term represents the  electrostatic interactions energy between monomers adsorbed on the
grafting plane and we neglect the volume interactions between the monomers. For a uniformly charged disc
of radius $R_g$ (which mimics a chain located randomly within such area) we have $\alpha = 16/(3 \pi)$,
while for a linearly stretched chain of length $2R_g$ this coefficient becomes $\alpha= (\log N_s +
\gamma_E-1)/2$.
The second term is purely entropic, giving the entropy of an uncharged two-dimensional chain the gyration radius of which is $R_g$.  It is expressed
in terms of the probability $P_{N_{s}}(R_{g})$
that a chain of $N_s$ monomers has a
gyration radius $R_g$, so that $(2 \pi)^{N_s} P_{N_{s}}(R_{g})$ gives the total number of such chain
configurations.
For two-dimensional chains and under the condition $R_g^2 \gg N_s
b^2 /6$, $P_{N_{s}}(R_{g})$ reads\cite{Fixman1962,BrilliantovSeidel2012}
\begin{equation}
\label{eq:P_g}  P_{N_{s}}(R_{g})=\frac{\pi^{3/2}e}{N_{s}}\exp\left(-\pi^2
\frac{R_{g}^2}{N_{s}b^2}\right).
\end{equation}
Note that in Eq.~(\ref{eq:surface}) interactions of the adsorbed part of the chain with the electric
field are neglected.
Minimising $F_{\rm surf}(R_{g}, N_s)$ with respect to the radius of gyration
$R_g$, we find that $R_g \sim N_s$ which justifies the application of Eq.~(\ref{eq:P_g}) and finally
the surface part of the free energy becomes
\begin{equation}
\label{eq:F_surf_min} \beta F_{\rm
surf}(N_{s})=N_{s}\left[p(q^2\tilde{l}_{B})^{2/3}-\log(2\pi)\right]+\log N_{s},
\end{equation}
with $p=3(\pi\alpha/2)^{2/3}$.
%

%
Collecting all contributions  of the chain free energy $F_{\rm tot}$, given by Eqs. (\ref{eq:F_bulk}),
(\ref{eq:F_self_bulk}),  (\ref{eq:spring}) and (\ref{eq:F_surf_min}) and minimising the sum with respect
to the number of desorbed monomers $N$ (recall that $N_s=N_0-N$ with $N_0 ={\rm const.}$) we obtain
the following equation for $N(E)$
\begin{equation}
\label{eq:Eq_for_N} \tilde{E} N + \tilde{l}_Bq^2 \log N = \tilde {\kappa} (\tilde{l}_0-N)^{\gamma} + Q \, ,
\end{equation}
where  $Q=[ p(\tilde{l}_Bq^2)^{2/3} - \log 2\pi]$. In the parameter $p=3(\pi\alpha/2)^{2/3}$  one has to insert the
value of $\alpha$ that corresponds to the chain configuration with the smallest total free energy.
Solving Eq.~(\ref{eq:Eq_for_N}) numerically, we obtain the equilibrium number of desorbed monomers of
the chain $N(E)$ which yields the force $f(E)$ generated by the applied electric field $E$ following the relation
\begin{equation}
\label{eq:F_of_N} f(E)    = \kappa (l_0-z_N)^{\gamma} \simeq  \frac{ \tilde{\kappa}}{\beta b} (
\tilde{l}_0-N)^{\gamma} \, .
\end{equation}
Theoretical results of force $f(E)$ as a function of applied field are shown in Figs. 4 and
5, where we compare them directly with MD data.

Interestingly, it turns out that the result obtained by the ambitious statistical physics approach presented above  is very close to the prediction of a
purely mechanical theory, based on the force balance between the electrostatic force acting on the bulk
part of the spring, $EeqN$ and the force from the spring $\kappa (l_0-bN)^{\gamma}$:
\begin{equation}
\label{eq:Mechan} EeqN^* \simeq \frac{ \tilde{\kappa}}{\beta b} (\tilde{l}_0-N^*)^{\gamma} \, ,
\end{equation}
where $N^*$ denotes the equilibrium number of desorbed monomers estimated by the mechanical theory.
For linear springs, i.e.,  at $\gamma =1$ Eq.~(\ref{eq:Mechan}) can be solved analytically and the simple mechanical theory yields
\begin{equation}
\label{eq:Mech_Theo} \tilde{f}(E) \simeq  \tilde{\kappa} \left(1-
\frac{\tilde{\kappa}}{\tilde{E}+\tilde{\kappa}} \right) \tilde{l}_0 \quad {\rm and } \quad N^* \simeq
\frac{\tilde{\kappa}\tilde{l}_0}{\tilde{E}+\tilde{\kappa}} \, ,
\end{equation}
where $\tilde{f}=\beta b f$. Note that Eq.~(\ref{eq:Mech_Theo}) shows both the explicit dependence of
generated force on electric field $E$ and bare spring length $l_0$. Because the deviations from the
simple mechanical theory are small, one can linearise Eq.~(\ref{eq:Eq_for_N}) around $N^*$. Considering
only terms linear in $\delta N = N - N^*$ we obtain the following corrections to generated force
\begin{eqnarray}
\label{eq:F_corect} \tilde{f}(E) &=&  \tilde{\kappa} (\tilde{l}_0-N^*-\delta N)^{\gamma} \\
\delta N &\simeq & \frac{ p(\tilde{l}_Bq^2)^{2/3} - \log 2\pi - \tilde{l}_Bq^2 \log N^*}{ \tilde{E}
+\gamma  \tilde{\kappa} \left( \tilde{E}N^*\!/\tilde{\kappa} \right)^{1-1/\gamma} } \, .
\end{eqnarray}
For the parameter range addressed here, the correction $\delta N$ is rather small, i.e., $\delta N \ll
N^*$. This circumstance explains the surprising accuracy of the purely mechanical theory. Upon
desorption of a finite number of monomers and transformation of the corresponding surface piece of the
chain into a bulk one several contributions to the total free energy compensate each other. Thus, finally the behaviour is dominated by the
purely mechanical balance between the two leading forces acting on the chain, i.e., on the one hand that by the electric field
and on the other hand that by the target body (spring).
%
 
  

For various applications, it would be convenient to have an approximate relation that expresses
explicitly the generated force in terms of applied field and geometric parameters of the system. For linear
springs, such a relation is given by Eq.~(\ref{eq:Mech_Theo}). For small fields $\tilde{E}$, one can
obtain a similar result  also for the general case of a non-linear force-deformation law. Using the approximate force balance
equation (\ref{eq:Mechan}) in the form
\begin{equation}
\label{eq:Mechan1} N^*=\tilde{l}_0 - \left( \frac{\tilde{E}N^*}{\tilde{\kappa}} \right)^{1/\gamma} \,
\end{equation}
one can iteratively solve it for $N^*$. With zero-order approximation $N^*=\tilde{l}_0$, we obtain
\begin{equation}
\label{eq:N_l0} N^*=\tilde{l}_0 - \left( \frac{\tilde{E} \tilde{l}_0 }{\tilde{\kappa}}
\right)^{1/\gamma} + \cdots \, .
\end{equation}
Applying the relation for the force $f \approx E qe N^*$, finally we get
\begin{equation}
\label{eq:f_l0} \tilde{f} \approx \tilde{E} \tilde{l}_0 - \tilde{E} \left( \frac{\tilde{E} \tilde{l}_0
}{\tilde{\kappa}} \right)^{1/\gamma} + \cdots \, ,
\end{equation}
which is supposed to be valid for small electric fields, $\tilde{E} \ll 1$. In Fig. 7, the above dependence is
compared with simulation data.
%

%
\section{Results and discussion}
%

%
In Figs.~4 -- 7 we compare the predictions of our theory with the results of MD
simulations. In the cases of linear and non-linear springs, see Fig.~4, as well as for Hertzian springs,
see Fig.~5, the agreement between theory and numerical data is rather good.
Moreover, we find that the simulation results obey quite well the purely mechanical theory. Evaluating correction terms within the statistical mechanics approach, it becomes evident that they effectively compensate each other for the parameter range addressed in the present study.
This particular feature may be important for applications because the simple theory yields useful estimates of the generated force as a function of applied field and
equilibrium spring length, see Eqs.~(\ref{eq:Mech_Theo}) and (\ref{eq:f_l0}). Interestingly,
the generated force depends almost linearly on the equilibrium spring length not only for linear
springs, which corresponds to coiled Gaussian chains, but also for non-linear, Hertzian springs (see Fig.~7)
which model a squeezed colloidal particles where the equilibrium length $l_0$ corresponds to the particle
diameter $d_c$.  From our previous simulations,\cite{BrilliantovSeidel2012} we can conclude that the response
force is almost independent of the total chain length $N_0$.

Another important observation refers to the impact of (undesired) electric current, caused by the
applied electric field, on the force generated. Simulations performed with Hertzian springs for
different current strengths  clearly demonstrate that the influence of current on
the magnitude of the force is almost negligible, see Fig. 6. This behaviour may be explained as follows: For small
currents, almost all counterions accumulate near the plane oppositely charged to the grafting plane, in
this way leaving the grafted chain unscreened. For large currents, counterions move rather fast and are
not able to screen efficiently. Note that weak impact of electric current on generated force is another
necessary condition for  the application of the simple mechanical theory.

\section{Conclusions}
In conclusion, we have studied theoretically and by means of MD simulations the force generation
(response force) by an external electric field in a system built up of a grafted polyelectrolyte chain
with the free end  linked to a target body. We study several force-deformation relations that specify
the mechanical properties of target bodies. In particular we analyse linear response which mimics coiled
polymer chains, non-linear (cubic)  response corresponding to semi-stretched chains and Hertzian
force-deformation relation that reflects the behaviour of squeezed colloidal particles. We observe  that
field-controlled mechanical force the magnitude of which depends on system parameters such as spring stiffness
and equilibrium spring length can be generated in a rather wide interval while the electric field is
varied only moderately. 
We develop a theory that  describes the phenomenon within a statistical
mechanics framework. Its predictions agree  quite well with simulation results. We demonstrate that due
to the mutual compensation of different contributions to the total free energy, a purely mechanical
theory based on simple force balance arguments may be accurately applied for the parameter range considered in the study. Since  the full theory can be
solved only numerically we also present a simplified approach which gives useful approximate relations for the dependence of 
generated force on various parameters.

To check the impact of undesired electric current across the setup, we performed simulations partially both at zero current
and at finite current strength: No noticeable impact of current on the generated force has been
detected.  Note that in the presence of additional salt ions (not considered here) the system boundaries
acting as electrodes become increasingly polarised which reduces the effective field acting on the
polyelectrolyte. We expect, however, that for small salt concentration and strong electric field this
effect would not qualitatively change the  phenomena addressed.

The systems studied may be treated  as prototypes of possible nano-devices where the force acting on a
target body can be regulated by external electric field. Nano-nippers or nano-vice designed on the basis
of the explored phenomenon and operated by external electric field, could be possibly used  to reversibly fix and
release nano-sized objects.


\newpage

\begin{center}
 Captions to illustrations
\end{center}
\begin{figure}[h]
\center{\includegraphics[width=1\linewidth]{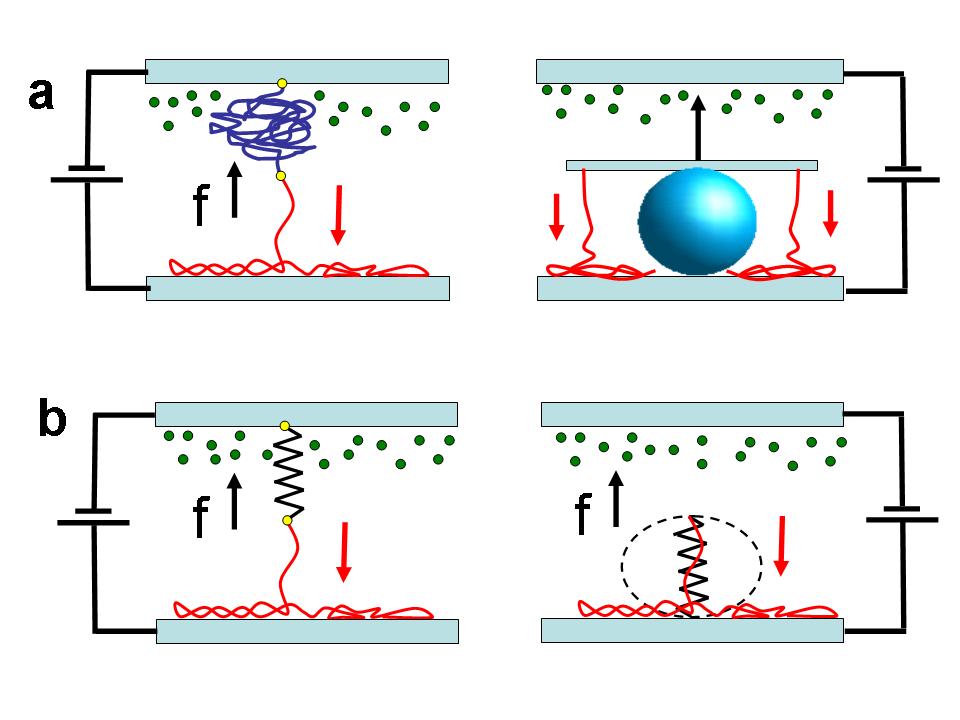}}
\caption{\sl  Two possible setups for field-regulated force generation.
   The force generated by an electric field that favours polyelectrolyte adsorption is
  indicated by red arrows down. The counteracting force $f$ caused by the deformation of the linker/body is shown by black arrows up. 
  The restoring force $f$ can depend both linearly or nonlinearly on deformation, including the Hertzian response of a compressed colloidal particle (right panel).
  The right panel also demonstrates the working principle of a  possible nano-vice: The target
  particle being fixed at sufficiently strong field will be released at vanishing or weak field.
 \textbf{(b)} The simulation setups modelling the "experimental" ones shown in the upper panel.  The free chain end is linked to a linear or
nonlinear spring which undergoes stretching with increasing field (left panel) or to a Hertzian spring which  endures
compression (right panel).}
\label{Fig:1}
\end{figure}

\begin{figure}[h]
\center{\includegraphics[width=1\linewidth]{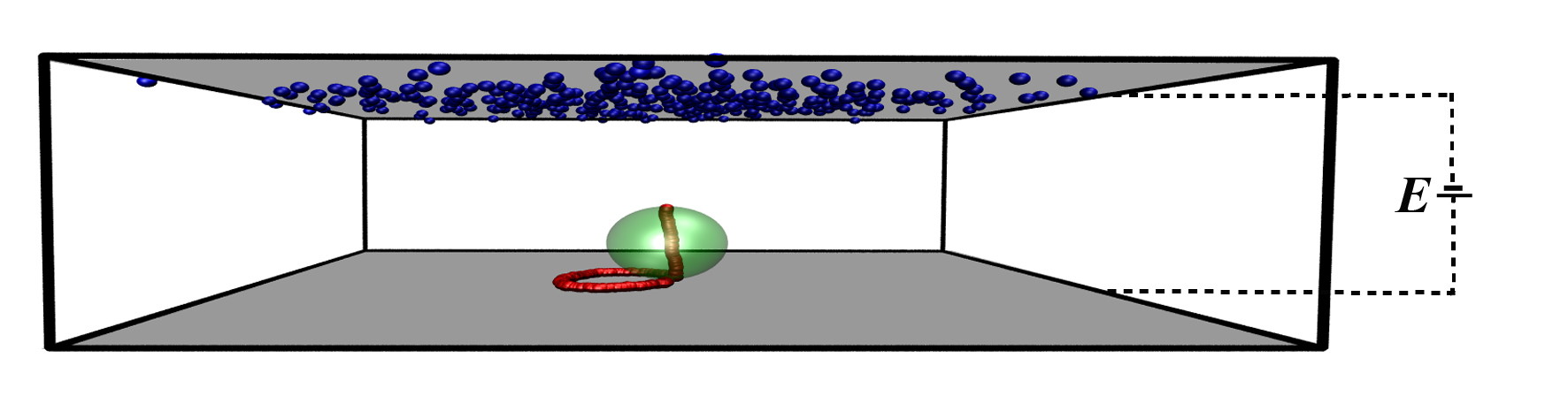}}
\caption{\sl  Typical simulation snapshot of a grafted polyelectrolyte ($N_0 = 320$) exposed
  to electrical field $E = 1 k_B T / eb$  perpendicular to the grafting plane and coupled
  to a deformable colloidal particle of diameter $l_0 = d_c= 80$ for vanishing current.
  In the simulations,  the action of the particle is modelled by the corresponding Hertzian force.}
\label{Fig:2}
\end{figure}

\begin{figure}[h]
\center{\includegraphics[width=1\linewidth]{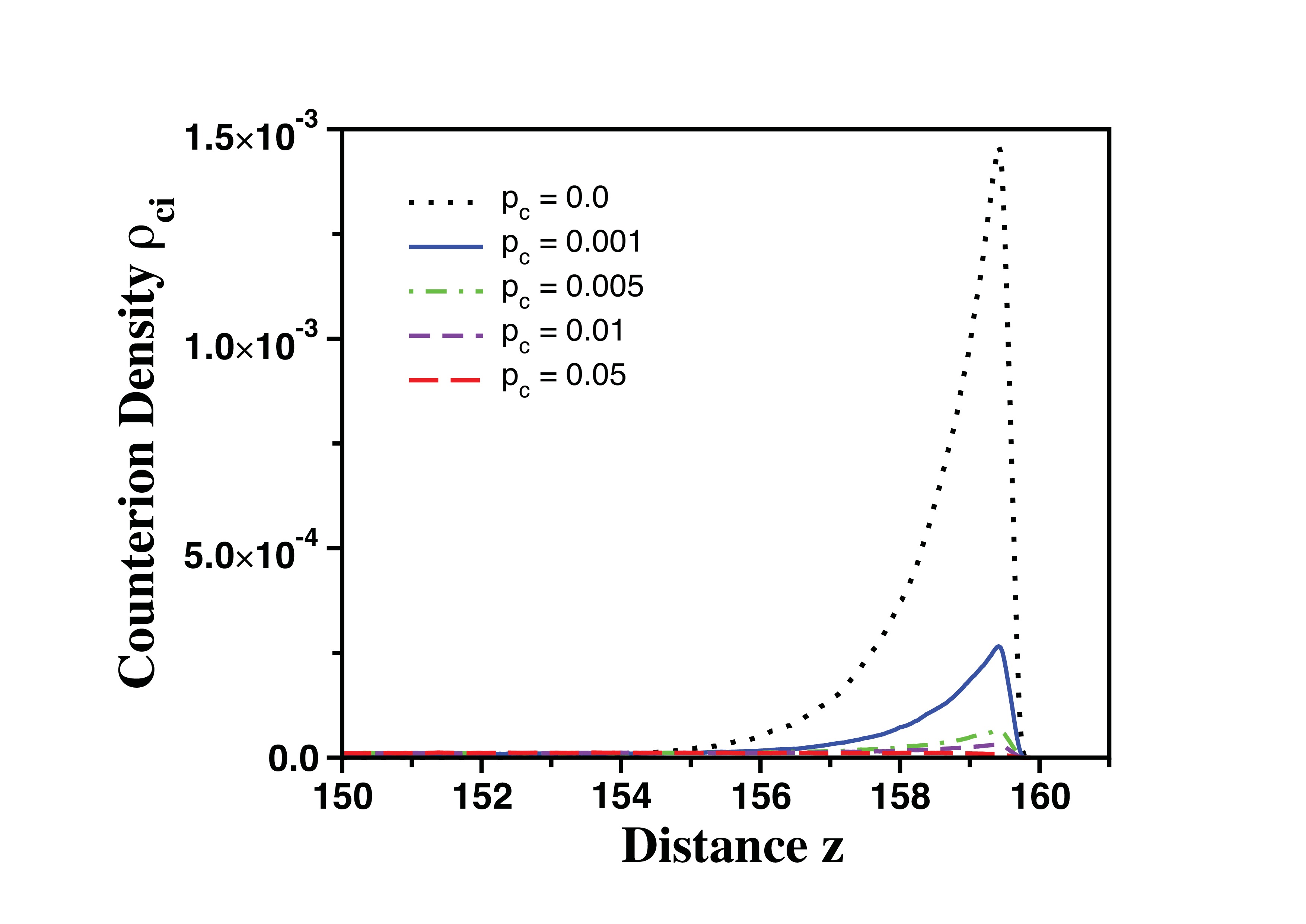}}
\caption{\sl  Average counterion density profile perpendicular to the grafting
  plane at $E = 1  k_B T / eb$  both at vanishing current, $j =a p_c$ with  $p_c = 0$
  ($a$ is a constant, see the text), and at finite current with  $p_c $ = 0.001, 0.005, 0.01 and  0.05.
  For vanishing current ($p_c = 0$), the density equals zero at $z < 152$ while for finite currents
  ($p_c>0$) it exhibits   a small but constant average value across the whole box,  except for a narrow
  layer close to the top boundary. Note that only a piece of the total box height of $L_z=160$ is
  plotted to show clearly the changes in the accumulation layer.}
\label{Fig:3}
\end{figure}

\begin{figure}[h]
\center{\includegraphics[width=1\linewidth]{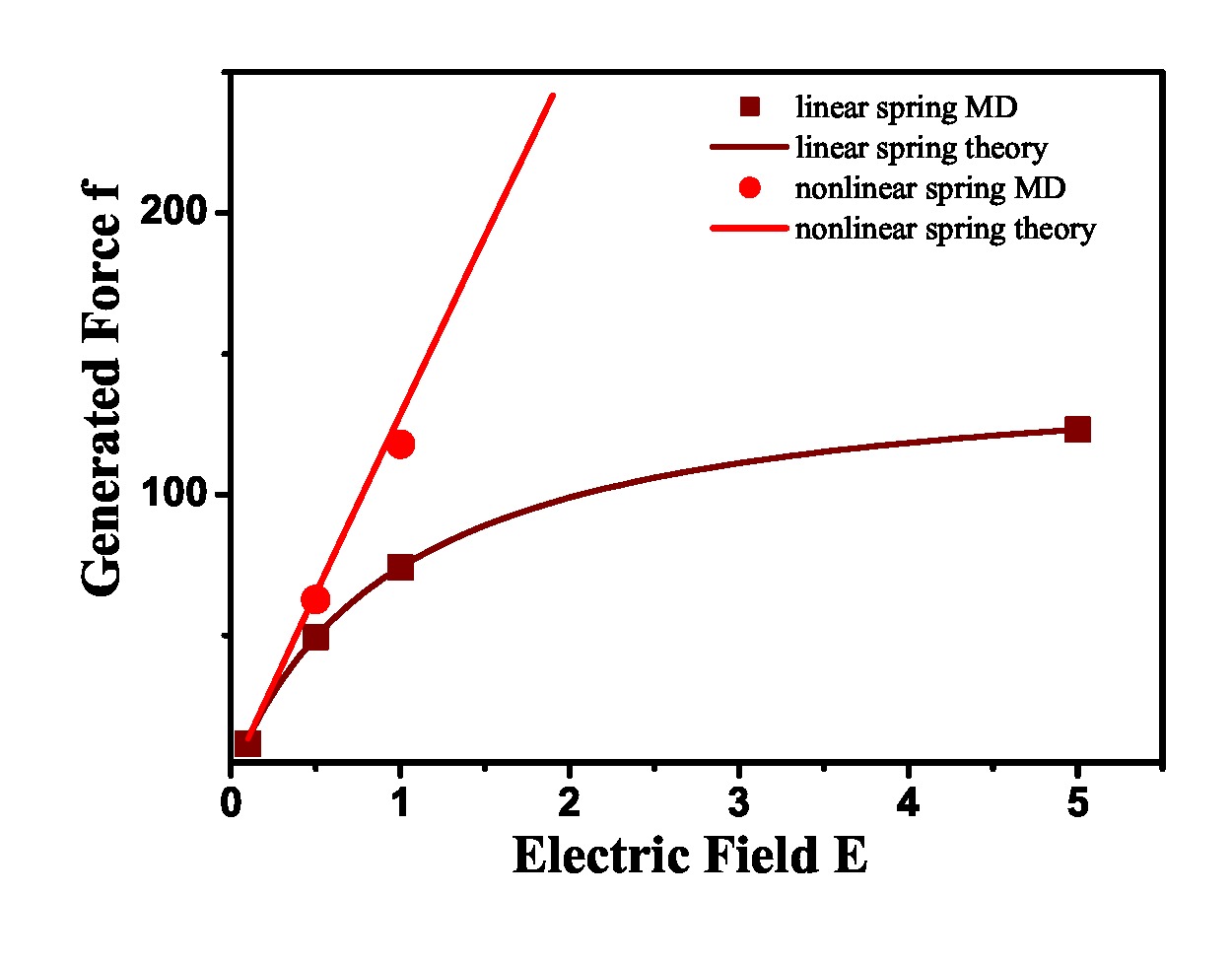}}
\caption{\sl  Average counterion density profile perpendicular to the grafting
  plane at $E = 1  k_B T / eb$  both at vanishing current, $j =a p_c$ with  $p_c = 0$
  ($a$ is a constant, see the text), and at finite current with  $p_c $ = 0.001, 0.005, 0.01 and  0.05.
  For vanishing current ($p_c = 0$), the density equals zero at $z < 152$ while for finite currents
  ($p_c>0$) it exhibits   a small but constant average value across the whole box,  except for a narrow
  layer close to the top boundary. Note that only a piece of the total box height of $L_z=160$ is
  plotted to show clearly the changes in the accumulation layer.}
\label{Fig:4}
\end{figure}

\begin{figure}[h]
\center{\includegraphics[width=1\linewidth]{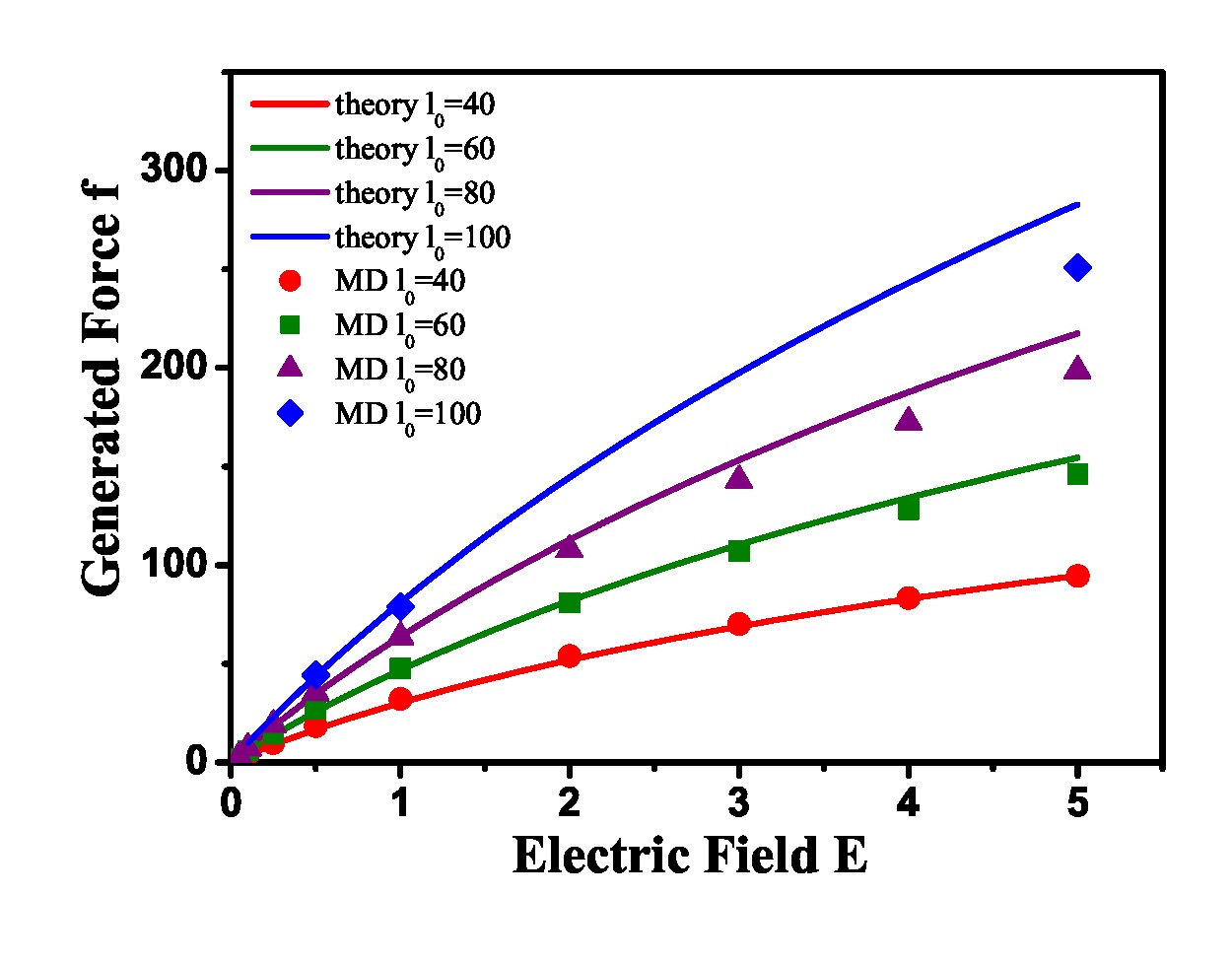}}
\caption{\sl  The dependence of generated force $\tilde{f}$ on applied electric field
$\tilde{E}$ for Hertzian springs with $\tilde{\kappa}=1$. Lines and symbols as in Fig.~4.}
\label{Fig:5}
\end{figure}

\begin{figure}[h]
\center{\includegraphics[width=1\linewidth]{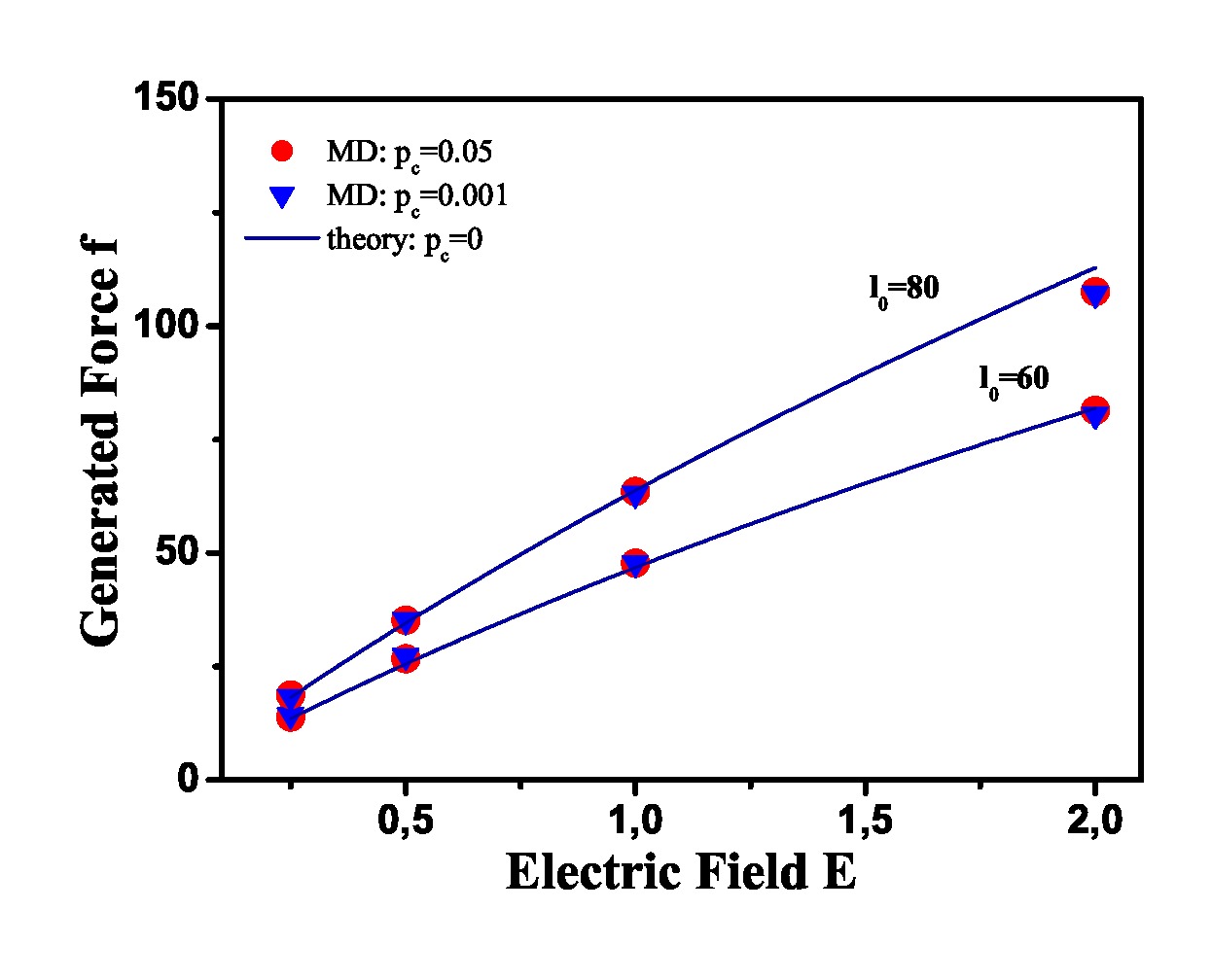}}
\caption{\sl  Generated force $\tilde{f}$ vs. applied electric field
$\tilde{E}$ for Hertzian springs with $\tilde{\kappa}=1$ at varying current strength. The current across 
the simulation box is proportional to $p_c$ (see the text).  Lines and symbols as in Fig.~4. 
Note that simulation data for zero current (see Fig.~5) are not shown because
symbols can not be distinguished from those of data at finite current.}
\label{Fig:6}
\end{figure}

\begin{figure}[h]
\center{\includegraphics[width=1\linewidth]{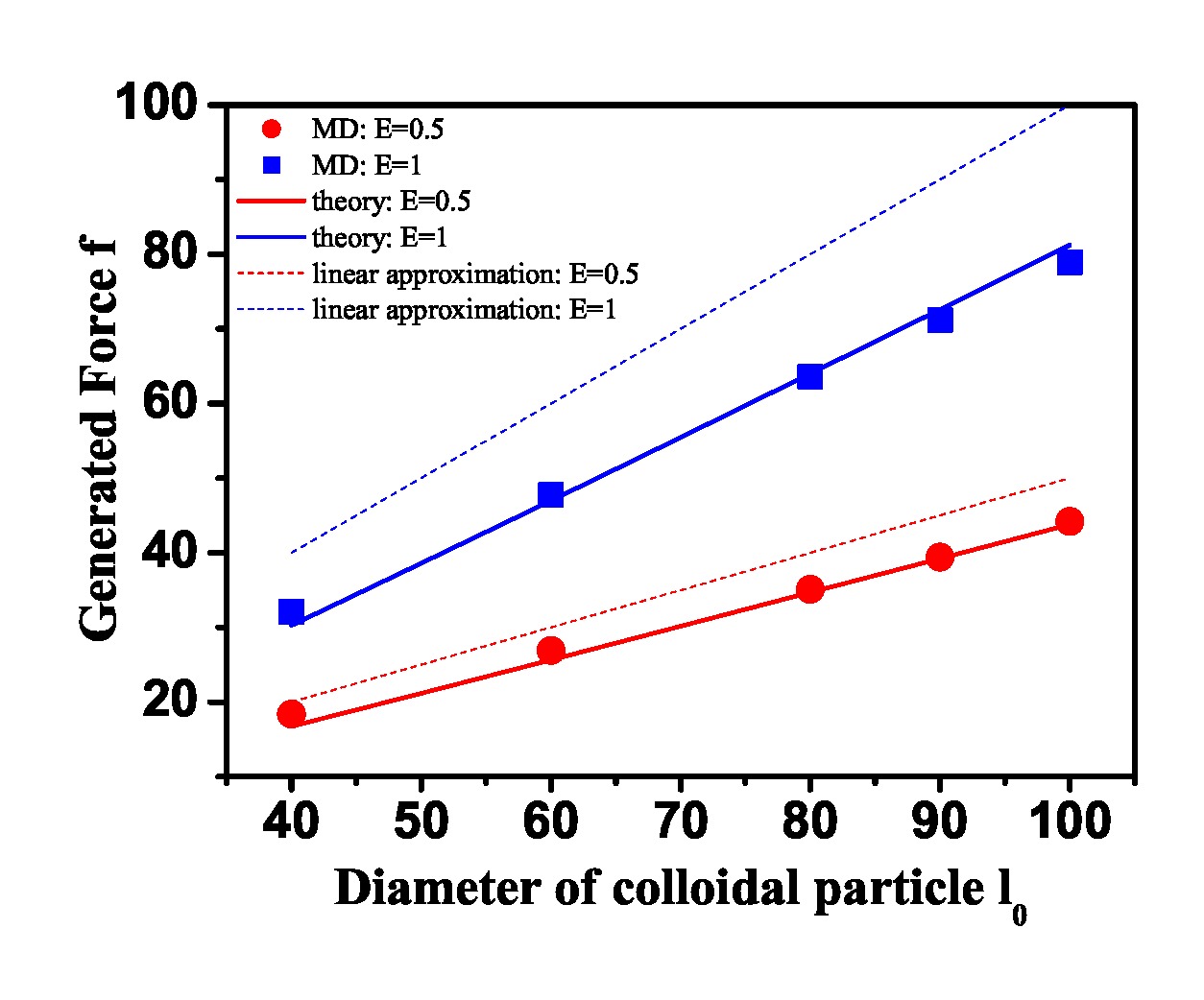}}
\caption{\sl  The dependence of generated force $\tilde{f}$ on equilibrium  length $l_0$
  for Hertzian  spring at zero electrical current. The strength of the spring  is $\tilde{\kappa}=1$.
  Note that a Hertzian spring of equilibrium length $l_0$ models a colloidal particle of diameter $d_c$
  which is in the range from about 3 nm to 70 nm according to our setting of length scale (see text).
  Solid lines and symbols as in Fig.~4, while dashed lines correspond to zero-order approximation in
  Eq.~(\ref{eq:f_l0}), that is,  for $\tilde{f} \approx \tilde{E}\tilde{l}_0$.}
\label{Fig:7}
\end{figure}


\begin{thebibliography}{00}
\bibitem{Muthu1987} M. Muthukumar {\sl}, J. Chem. Phys.  86(1987) 7239.

\bibitem{Bajpai1997} A. K. Bajpai {\sl }, Prog. Polym. Sci.  22(1997) 523.

\bibitem{Borisov1994} O. V. Borisov, E. B. Zhulina, and T. M. Birshtein {\sl }, J. Phys. II France  4(1994) 913.

\bibitem{Boru98} I. Borukhov and  D. Andelman and H. Orland {\sl }, Macromolecules  31(1998) 1665.

\bibitem{Joanny98} X. Chatellier, and J.-F. Joanny {\sl}, Phys. Rev. E 57(1998) 6923.

\bibitem{Muthu2004} M. Muthukumar {\sl}, J. Chem. Phys. 120(2004) 9343.

\bibitem{Dobry2000} A. V. Dobrynin,  and A. Deshkovski, and M. Rubinstein {\sl}, Phys. Rev. Lett.  84(2000) 3101.

\bibitem{Dobry2001} A. V. Dobrynin,  and A. Deshkovski, and M. Rubinstein {\sl}, Macromolecules  34(2001) 3421.

\bibitem{Borisov2001} O. V. Borisov, and  F. A. M. Leermakers, and  G. J. Fleer, and E. B. Zhulina {\sl}, J. Chem. Phys. 114(2001) 7700.

\bibitem{Netz2003} R.R. Netz {\sl}, Phys. Rev. Lett. 90(2003) 128104.

\bibitem{FriedsamGaubNetz2005} C. Friedsam, and H. E. Gaub, and R. R. Netz {\sl}, Europhys. Lett. 72(2005) 844.

\bibitem{BrilliantovSeidel2012} N. V. Brilliantov, and C. Seidel {\sl}, Europhys. Lett. 97(2012) 28006.

\bibitem{Netz2003a} R. R. Netz {\sl}, J. Phys. Chem. B 107(2003) 8208.

\bibitem{Borisov2003} O. V. Borisov, and A. B. Boulakh, and E. B. Zhulina {\sl}, Eur. Phys. J. E 12(2003) 543.

\bibitem{Podg1993} P. Podgornik, and B. Jonsson {\sl}, Europhys. Lett. 24(1993) 501.

\bibitem{Podg1995} P. Podgornik and T. Akesson, and B. Jonsson {\sl}, J. Chem. Phys. 102(1995) 9423.

\bibitem{Podg2006} P. Podgornik, and M. Licer {\sl}, Curr. Op. Coll. Interf. Sci. 11(2006) 273.

\bibitem{Hisao:2009} H. Kuninaka, and H. Hayakawa {\sl}, Phys. Rev. E 79(2009) 031309.

\bibitem{Hisao:2010} K. Saitoh, and A. Bodrova, and H. Hayakawa, and  N. V. Brilliantov {\sl}, Phys. Rev. Lett. 105(2010) 238001.

\bibitem{Netz2008} S. Fischer, and A. Naji, and  R. R. Netz {\sl}, Phys. Rev. Lett. 101(2008) 176103.

\bibitem{CSA00} F. S. Csajka, and C. Seidel {\sl}, Macromolecules 33(2000) 2728.

\bibitem{KUM05} N. A. Kumar, and C. Seidel {\sl}, Macromolecules 38(2005) 9341.

\bibitem{Winkler98} R. G. Winkler, and M. Gold, and P. Reineker {\sl}, Phys. Rev. Lett. 80(1998) 3731.

\bibitem{Gole99} R. Golestanian, and T. B. Kardar and Liverpool {\sl}, Phys. Rev. Lett. 82(1999) 4456.

\bibitem{Pincus1998} H. Schiessel, and P. Pincus {\sl}, Macromolecules 31(1998) 7953.

\bibitem{MickaHolm1999} U. Micka, and  C. Holm, and K. Kremer {\sl}, Langmuir 15(1999) 4033.

\bibitem{Diehl96} A. Diehl, and M. Barbosa, and   Y. Levin {\sl}, Phys. Rev. E 54(1996) 6516.

\bibitem{Naji:2005} A. Naji, and R. R. Netz {\sl}, Phys. Rev. E 95(2005) 185703.

\bibitem{Fixman1962} M. Fixman {\sl}, J. Chem. Phys. 26(1962) 185703.

\bibitem{Viovy2000} J.-L. Viovy {\sl}, Rev. Mod. Phys. 72(2000) 813.

\bibitem{Klep2007} K. Kleparnik, and P. Bocek {\sl}, Chem. Rev. 107(2007) 5279.

\bibitem{Friedsam2005} C. Friedsam, and H. E. Gaub, and R. R. Netz {\sl}, Europhys. Lett. 72(2005) 844.


\end{thebibliography}
\end{document}